\documentclass[12pt,a4paper]{article}

\usepackage{amsmath}
\usepackage{scicite}
\usepackage{times}
\usepackage{graphicx}
\usepackage{tabularx}
\usepackage{booktabs}

\newenvironment{sciabstract}{%
\begin{quote} \bf}
{\end{quote}}

\title{Limited containment options of COVID-19 outbreak\\ revealed by regional agent-based simulations\\ for South Africa}
\author
{
\begin{minipage}{1\textwidth}
Andreas Bossert\,\textsuperscript{1,7}, 
Moritz Kersting\,\textsuperscript{2,7}, 
Marc Timme\,\textsuperscript{3,6}$^\ast$, 
Malte Schr{\"o}der\,\textsuperscript{3}, 
Azza Feki\,\textsuperscript{4,7}, 
Justin Coetzee\textsuperscript{5},
and Jan Schl{\"u}ter\,\textsuperscript {6,7}$^\ast$\\
\\
\normalsize{$^{1}$Center of Methods in Social Sciences, Department of Social Sciences, Georg August University of Göttingen, Goßlerstraße 19, 37073 Göttingen, Germany}\\
\normalsize{$^{2}$Chair of Statistics, Department of Economics, Georg August University of Göttingen, Humboldtallee 3, 37073 Göttingen, Germany}\\
\normalsize{$^{3}$Chair for Network Dynamics, Institute for Theoretical Physics and Center for Advancing Electronics Dresden (cfaed), Technical University of Dresden, Helmholtzstr. 18, 01069 Dresden, Germany}\\
\normalsize{$^{4}$Chair of Software Engineering, Faculty of Natural Sciences and Technology, Hochschule für angewandte Wissenschaft und Kunst, Von-
Ossietzky-Straße 99, 37085 Göttingen, Germany}\\
\normalsize{$^{5}$GoMetro, 10 Church Street, Durbanville, Cape Town, South Africa, 7550}\\
\normalsize{$^{6}$Institute for the Dynamics of Complex Systems, Faculty of Physics, Georg August University of Göttingen, Friedrich-Hund-Platz 1, 37077 Göttingen, Germany}\\
\normalsize{$^{7}$Next Generation Mobility (NGM), Department of Dynamics of Complex Fluids, Max-Planck-Institute for Dynamics and Self-Organization, Am Fassberg 17, 37077 G\"ottingen, Germany}\\[0.5em]
\normalsize{$^\ast$To whom correspondence should be addressed: jan.schlueter@ds.mpg.de (JS) or marc.timme@tu-dresden.de (MT)}
\end{minipage}
}

\date{}

\begin{document} 


\baselineskip24pt


\maketitle

\vspace{1.5cm}
{\textit{\large{\underline{One Sentence Summary}:
Regional agent-based scenario simulations of the COVID-19 outbreak in South Africa 
%Simulating the local outbreak of COVID-19 
%by combining large scale socio-economic and geographic data 
%with a detailed regional mobility model including widespread use of paratransit
%for a municipality in South Africa 
suggest that countermeasures in many countries of the Global South are required to be especially rapid and severe
%only most severe and immediate countermeasures may 
to not overload the capacity set by locally available intensive care units.
%highlighting boundary conditions for constraining pandemics across countries of the Global South.
}}}

\newpage


\begin{sciabstract}
 COVID-19 has spread from China across Europe and the United States and has become a global pandemic. In countries of the Global South, due to often weaker socioeconomic options and health care systems, effective local countermeasures remain debated. We combine large-scale socioeconomic and traffic survey data with detailed agent-based simulations of local transportation to analyze COVID-19 spreading in a regional model for the Nelson Mandela Bay Municipality in South Africa under a range of countermeasure scenarios. The simulations indicate that any realistic containment strategy, including those similar to the one ongoing in South Africa, may yield a manifold overload of available intensive care units (ICUs). Only immediate and the most severe countermeasures, up to a complete lock-down that essentially inhibits all  joint human activities, can contain the epidemic effectively. 
 %The shared  paratransit transport services significantly enhances this effect. 
 %This effect is significantly enhanced by the use of shared paratransit services as the main mode of transport.
 As South Africa exhibits rather favorable conditions compared to many other countries of the Global South, our findings constitute rough conservative estimates and may support identifying strategies towards containing COVID-19 as well as any major future pandemics in these countries.
\end{sciabstract}

% In setting up this template for *Science* papers, we've used both
% the \section* command and the \paragraph* command for topical
% divisions.  Which you use will of course depend on the type of paper
% you're writing.  Review Articles tend to have displayed headings, for
% which \section* is more appropriate; Research Articles, when they have
% formal topical divisions at all, tend to signal them with bold text
% that runs into the paragraph, for which \paragraph* is the right
% choice.  Either way, use the asterisk (*) modifier, as shown, to

\newpage

%\section*{Introduction}

The first cases of the corona virus disease COVID-19 were confirmed on December 29, 2019 in China \cite{Li2020}. Due to high contagiousness \cite{sanche2020novel}, the disease has spread rapidly \cite{hufnagel2004forecast, brockmann2013hidden} and the World Health Organization (WHO) has since confirmed \cite{WHO2.2020} about 1.6 million cases worldwide, with its vast majority as of this writing (April 11, 2020) in the United States and many countries in Europe. Most countries of the Global South are affected as well, often with reported case numbers just beginning to surge \cite{WHO.2020}. However, the currently reported low numbers, e.g. about 9000 cases in all of Africa, may be biased as only few  individuals have been tested, see \cite{sacoronavirus} for the example of South Africa, and also \cite{AustriaUndersampling}. Furthermore, such countries operate under vastly different socio-economic conditions, with larger social inequality, distinct transportation options and weaker health care systems compared to those of most countries in the global north \cite{hunt2014exponential, Bannon2003NaturalRA, gilbert2020preparedness}, such that core characteristics of potential COVID-19 spreading dynamics and thus the effectiveness of specific countermeasures remain largely unknown for countries of the Global South. 

The transmission of COVID-19 is currently thought to occur through direct inter-person droplet-based infections through coughing and sneezing, with possible additional infection paths through aerosols and via contaminated surfaces. Symptoms often occur only after an incubation period of several days, with many infected responding entirely asymptomatic. 
%Also symptom-free humans are contagious such that tracing and estimating spreading patterns is challenged \cite{RKI.2020}. 
Contagion may also occur via such symptom-free carriers, posing a challenge for tracing and estimating spreading patterns \cite{RKI.2020}. 
Transmission predominantly occurs through interactions while being at home, being at work, using shared transportation modes or during group-based leisure activities. In many countries of the Global South, most of these social activities occur under
conditions drastically different from those in, for example, Europe or the U.S..
For instance, these countries feature, on average, more people per household, higher unemployment rates, more manual and lower payed work \cite{un2017household, ilo2020employment} and much of publicly available transportation services used by the middle and low-income population are offered as paratransit shared-mobility services \cite{hilling2003transport, Simons2020MinibusSouthAfrica}. Moreover, public health care services are different and often less well equipped than in countries of the Global North \cite{tandon2000measuring, gilbert2020preparedness}.

As a paradigmatic region offering large-scale data availability, 
we consider the Nelson Mandela Bay Municipality (NMBM) in South Africa to study scenarios of COVID-19 spreading dynamics and the impact of countermeasures. We combine socio-economic and travel survey data from more than 100,000 people \cite{JOUBERT20181012}, about 10\% of the local population, %including 
based on employment status, household size, age group and income level together with a detailed 24-hour travel diary component integrated into an agent-based traffic simulation (see Materials and Methods). The resulting contact network forms the basis for extended Susceptible-Infected-Recovered (SIR) model dynamics with parameters adapted to COVID-19 \cite{mueller2020}. 
Such agent-based simulations capture the inhomogeneity among the agents and are capable of modeling intricate nonlinear dynamic relationships between them, including a spreading rate that depends on the individual agent's detailed activities, their modes of transportation used, and their distance to each other \cite{BenArieh2014}. 
%%%(risk group)

To evaluate the impact of various policy measures on the course of the disease, 
we
% classify the population into a \textit{high risk group} and a \textit{normal risk group} and 
%distinguish 
systematically compare several scenarios by varying simulation parameters accordingly (see Materials and Methods). First, a \textit{baseline} scenario without any countermeasures; second, a \textit{default} scenario in line with the current measures implemented in South Africa as of early April 2020 \cite{businesslive}. These include the shutdown of childcare and educational institutions, the prohibition of leisure activities of any kind and cutting shopping activity options by about 70\,\%. Moreover, work related and ``other'' activities (for example trips to health care facilities and visits to public institutions) and travelling as passenger in a car are reduced by 80\,\% and formal and informal public transport is reduced by 30\,\%.  Third, as a harsher variation of the lockdown that may be achievable in principle, we study the effect of a \textit{realistic lockdown} scenario, increasing the restriction of activities related to work, shopping, leisure, and ``other'' by 90\,\% while childcare, educational activities as well as formal and informal public transport are completely shut down. Finally, we consider a theoretical \textit{complete lockdown} where all travel and outside activities are prohibited. 

All countermeasures come into effect 7 days after an initial infection of 100 people. 
%(as of April 9, 2020, 30 cases have been confirmed in the Nelson Mandela Bay Municipality, however, the number of true cases is likely much higher \cite{nmbm_covid_cases_news, AustriaUndersampling}). 
We furthermore investigate additional simulations that start with the (currently enacted) default measures and introduce the realistic lockdown scenario after a number of days. % that we vary.
%To identify the influence of paratransit mobility services, we additionally consider a variation of each of the scenarios where the infectiousness among people in paratransit vehicles and when waiting for such transportation is set to zero. For each of the four scenarios, a comparison to this variant directly reveals 
%how many people are infected through interpersonal contacts made by the sharing of public transport (paratransit) vehicles.
%the impact of interpersonal contacts made by the sharing of public transport (paratransit) vehicles on the speed and total magnitude of the outbreak.
%rather than in the absence of risk when still making the same trips and the same times.
The results presented below may inform and complement the ongoing discussion around tightening, loosening, introducing or repealing certain countermeasures.

Direct model simulations in the four scenarios (Figure~1) reveal that a large fraction of the population becomes infected, not only for the baseline scenario without countermeasures (as expected), but also with the default countermeasures active as of this writing (April 11 2020), see Figure panels 1A and 1B. Even if scenarios of harsh but realistically possible or complete lockdown were enacted only seven days after reaching 100 infected people, the outbreaks would still become macroscopic, with tens of thousands of infected people (Figure 1C and 1D) in the NMBM alone. Across all scenarios enacted, outbreak dynamics also causes overload of the approximately ICU beds estimated as baseline for NMBM (Figure~1E,F). 
As a precondition for successful containment, lockdown needs to be implemented sufficiently early on (in particular less than 7 days after the first 100 infected, \textit{not} the first 100 confirmed infected).
%
 
%
%
%%

% Figure 2 illustrates the dynamics in an effective state space of the outbreak, displaying newly reported cases \textit{vs.} the total integrated number of cases reported so far. It highlights the overall effectiveness of countermeasures independent of the absolute time it sets in and thereby makes counteractions directly comparable.

%Figure 2 illustrates the effectiveness of the countermeasures.
In all four scenarios, the outbreak strongly overloads the available intensive care unit (ICU) beds, with ten- to 100-fold overload in default and baseline scenario, respectively (see Figure 2A). 
Out of the total $267$ ICU beds available in (public and private) hospitals in the whole Eastern Cape Province (data from 2008, \cite{SAMJ6415}), only about $50$ would be available for critical cases from the NMBM (scaled proportional to the relative population counts in 2011 \cite{wiki_eastern_cap, wiki_mnmb}). 
%In the whole whole Eastern Cape Province, $267$ ICU beds are available in (public and private) hospitals. Even if all of these beds were available for the Nelson Mandela Bay Muncipality, this capacity is widely exceeded during the peak of the outbreak in all scenarios except severe lockdown.
We additionally quantify the sustained pressure on the health care system by computing the cumulative overload $\lambda$ of the health system as the total number of person days critical patients go without intensive care (Figure 2B), 
$$\lambda = \int \Theta(c(t) - c_\mathrm{max}) \left[c(t) - c_\mathrm{max}\right] \, \mathrm{d}t$$ where $\Theta(\cdot)$ denotes the Heaviside step function, $\Theta(x)=1$ for $x\geq0$ and $\Theta(x)=0$ otherwise,  $c(t)$ the number of critical cases at time $t$ and $c_\mathrm{max}$ the available ICU capacity. The integral is taken over the entire time of simulation, 64 days after the 100 people are infected in NMBM. %Whereas the peak value measures the maximal instantaneous overload, the cumulative overload $\lambda$ quantifies the sustained overload of the private and public health services, even when the number of cases only barely exceed the capacity but do so for a long time.
This cumulative measure $\lambda$ thus quantifies the total long term overload of intensive care health services (and may be large even if the peak overload is small). 
%, even when the number of cases just barely exceed the capacity but does so for a long time.

The potential severity of the outbreak in NMBM
%(and likely many other regions in the Global South) 
even with strong countermeasures, is especially evident when compared to that of the United Kingdom (UK) (Figure~2C). In the UK, ICU capacity would be exceeded by a factor of about 35 without any countermeasures, by  factor of about 2 with social distancing (similar to default in our simulations) and not at all with measures currently enacted \cite{Ferguson.2020}. In our NMBM scenario simulations, ICU capacity would be exceeded by a factor of more than 200 without any countermeasures, by factor of about 20 with default measures, by a factor of 5 with realistic lockdown and still by a factor of 3 even with complete lockdown. These numbers underline the strong fragility of the health care services by example of the NMBM. We expect similar orders of magnitude of capacity exceedance throughout South Africa and in many countries of the Global South. 

%Overall, the capacity of the available ICUs is exceeded by a factor of 36 in the baseline case without any countermeasures. Even severe realistic countermeasures will result in ...\% more critical cases during the peak of the outbreak than available ICU capacity. Comparing these results to predictions for the UK of about 35-fold overload without any countermeasures \cite{Ferguson.2020} demonstrates how much more fragile the health services in South Africa and countries of the Global South may be (Fig. 2C).

%%%
%Importantly, this is not only due to the less equipped health system but also due to the other local conditions. In particular, the role local paratransit plays in speeding up the spread of the disease becomes evident from simulations with zero infections during public transit activities (compare above). In this setting, the peak severity of the outbreak is reduced by up to 25\%. (Figure [])
%OR
%In a setting where infections cannot occur during usage of the paratransit services, the impact of the outbreak in the currently enacted default setting is reduced by more than 90\% (Fig. 2, light colored bars), underlining the large influence of local paratransit mobility for speeding up disease spreading (see also Supplementary Material, Table S2). 
%%%

Therefore, all realistic scenarios, including the default scenario currently enacted in South Africa, hardly seem capable of containing the COVID-19 pandemic without substantially overloading ICU availability. Our results suggest that such overload %would even be strong 
could even be expected if all the $267$ ICU beds in both private and public hospitals within the entire Eastern Cape Province would be exclusively available for critical COVID-19 patients in the Nelson Mandela Bay Municipality only.

Realistic or complete lockdown offer options 
of greatly reducing the pressure on the 
health care system if enacted rapidly enough. However, economic boundary conditions and in particular the large inequality in South Africa -- and similarly in many other countries of the Global South -- pose additional problems.  Thus, in our opinion, in particular complete lockdown seems hardly enforceable and would appear unsustainable especially for a large share of people with lower income, as already weaker containment will likely lead to severe economic consequences in addition to disease related deaths \cite{statssa2019a, statssa2019c}. 

How can the number of critical cases be confined given that the default scenario is already enacted? On April 9th, 2020, the number of cases reported in NMBM was 30 \cite{nmbm_covid_cases_news}. As the number of reported cases seem to systematically underestimate their actual number by at least a factor of three, as predicted for Austria on April 10, 2020 \cite{AustriaUndersampling}, we estimate that on 9th of April, at least %about 
100 cases existed in NMBM and take that as our initial condition. 
Starting simulations with the default scenario active, %on that day 
we find that critical cases are likely to vastly exceed the available ICU capacity (Figure 3A) within two months. 
In contrast, introducing the realistic lockdown scenario immediately, i.e. starting April 13th, might support a successful confinement of the number of critical COVID-19 patients in NMBM to manageable numbers (Figure 3B). In this scenario, the simulations suggest that there is about 80\% likelihood that ICU capacity becomes overloaded at some time and that ,if overload occurs, it will be relatively mild (factor of 2-3 of the capacity, in contrast to factor of about 30 when keeping the default measures as they are).

We remark that the success of such lockdown countermeasures crucially relies on several restrictions and our simulations likely under- rather than overestimate future case numbers. First in our simulations, activities are immediately reduced to the set low values (complete shutdown of public transport, child care and educational facilities, 90\% reduction of all other activities) and all inhabitants fully comply with these restrictions. Second, our simulations are based on estimated parameters that thereby come with uncertainties, and stochastic dynamics may create large deviations from the predicted values, in particular also earlier growth and larger total number of critical patients, not last due to an exponential growth of the outbreak in its initial phase without severe lockdown (compare \cite{maier2020effective}).
Third, our estimates assume that all ICU beds would be available exclusively for COVID-19 patients during the entire time of the outbreak. Those and other constraints call for a more conservative, especially earlier introduction of realistic countermeasures.

Across all scenarios studied, the results thus indicate that it may be hard to enact realistic, socially and economically feasible countermeasures without exceeding ICU capacity and that more drastic measures beyond the current default are rapidly needed. Finally, it seems reasonable to assume that the consequences of countermeasures would be qualitatively the same across South Africa as well as many countries of the Global South.
% Your references go at the end of the main text, and before the
% figures.  For this document we've used BibTeX, the .bib file
% scibib.bib, and the .bst file Science.bst.  The package scicite.sty
% was included to format the reference numbers according to *Science*
% style.

%BibTeX users: After compilation, comment out the following two lines and paste in
% the generated .bbl file. 

\nocite{Horni.2016,JOUBERT20181012, neumann2015, bischoff2017, neumann2015, gtfs2011, algoabus, mueller2020, Kermack.1927, Keeling.2008, Anderson.2010, Roche.2011, mueller2020}

\bibliographystyle{Science}
\bibliography{references.bib}

\section*{Acknowledgments}
MS and MT acknowledge support by the German National Science Foundation (DFG) and the Saxonian State Ministry for Higher Education, Research and the Arts through the Center for Advancing Electronics Dresden (cfaed). Author contributions:  MT, MS and JS conceived and designed research. AB and MK designed, set up and adapted the simulation software and ran simulations, supervised by JS. AB, MK, BS and JS evaluated the simulation data. MS and MT provided theoretical background and advised on data analysis and data presentation. AB, MK, MT, MS and JS wrote the basic version of the manuscript. JC provided local data for NMBM and advised on conditions for mobility simulations. All authors interpreted the results and contributed to revising and editing the manuscript. Competing interests: None declared.
%Here you should list the contents of your Supplementary Materials -- below is an example. 
%You should include a list of Supplementary figures, Tables, and any references that appear only in the SM. 
%Note that the reference numbering continues from the main text to the SM.
% In the example below, Refs. 4-10 were cited only in the SM.     

\section*{Supplementary materials}
Materials and Methods\\
%Supplementary Text\\
Table S1 through S4\\
%Figure S1\\
References \textit{(30-38)}

% For your review copy (i.e., the file you initially send in for
% evaluation), you can use the {figure} environment and the
% \includegraphics command to stream your figures into the text, placing
% all figures at the end.  For the final, revised manuscript for
% acceptance and production, however, PostScript or other graphics
% should not be streamed into your compliled file.  Instead, set
% captions as simple paragraphs (with a \noindent tag), setting them
% off from the rest of the text with a \clearpage as shown  below, and
% submit figures as separate files according to the Art Department's
% instructions.

\clearpage

% Legende: 
%- oben links, baseline
%- oben mitte, realistic lockdown
%- oben rechts, critical aus allen vier szenarien (gleiche farbe)
%- unten links, default config
%- unten mitte, complete lockdown
%- unten rechts, zoom von oben rechts

{
\centering
\includegraphics[width=\textwidth]{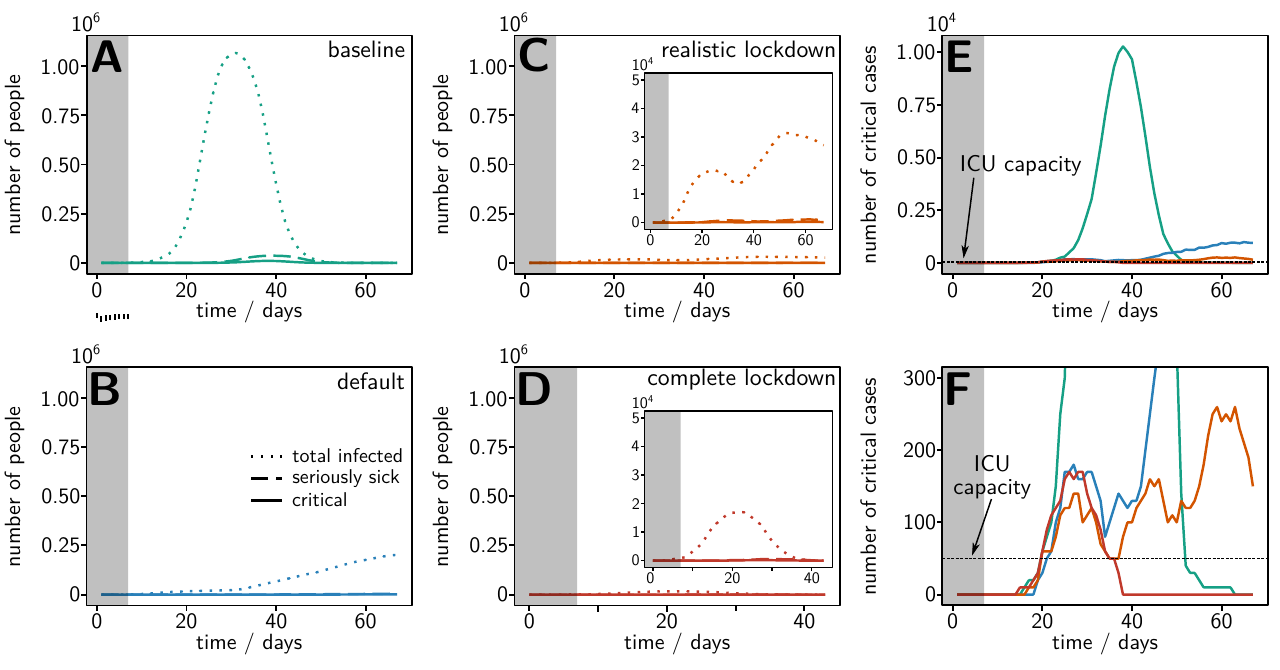}
}

\noindent {\bf Fig. 1. Distinct outbreak scenarios of COVID-19}. Dynamics for the Nelson Mandela Bay Municipality, Eastern Cape, South Africa, with a total population of 1.15 million people. (A-D) Evolution of the number of infected (dotted lines), seriously sick (dashed) and critical (solid) patients in four different scenarios: (A) baseline without any countermeasures, (B) with default countermeasures in line with current conditions in South Africa, (C) with harsh realistic countermeasures shutting down 90\% of all activities, and (D) with complete lockdown prohibiting any movement and group activities (infections within a household are still possible, see Supplementary Material for details). All countermeasures are initiated 7 days (grey shading) after the an initial infection of 100 people. 
(E) and (F) Number of critical cases requiring intensive care compared to the available ICU capacity (horizontal  dashed line) for all four countermeasure scenarios. None of the scenarios reduces the number of critical cases sufficiently to guarantee intensive care (ICU) treatment for all critical patients. %Note that the countermeasures currently in place started before 100 cases in the Nelson Mandela Bay Municipality were reported, see also Fig.~3.

\newpage

{
\centering
\includegraphics[width=\textwidth]{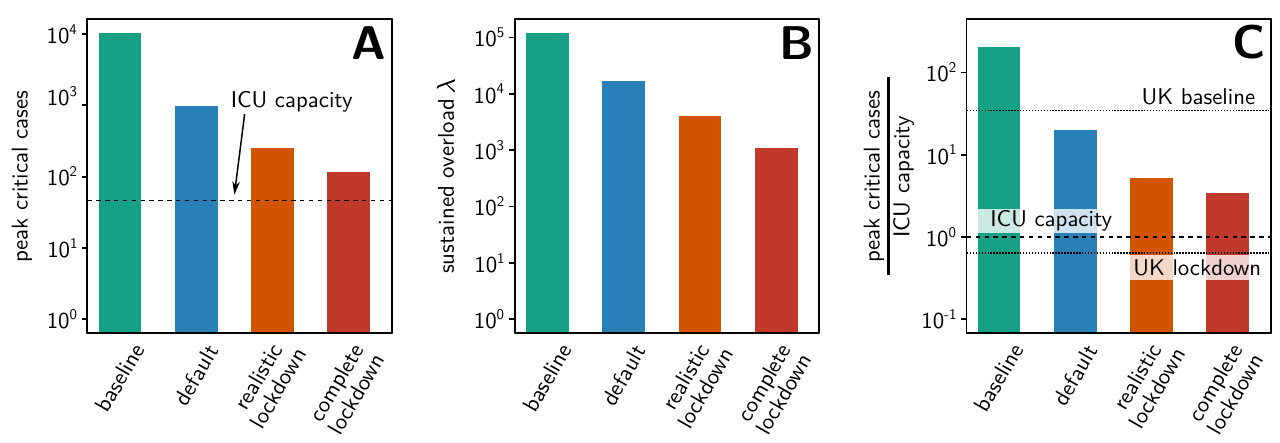}
}

\noindent {\bf Fig. 2. Overload of the health care system.} 
(A) Maximum number of concurrent critical patients during the peak of the outbreak. The available ICU capacity (dashed line) is widely exceeded in all scenarios. 
(B) The measure $\lambda$ for the sustained overload, counting the total number of person days that critical patients are without intensive care. 
(C) Relative exceedance of the ICU capacity at the peak of the outbreak. The potential consequences of COVID-19 in countries of the Global South becomes clear from the comparison to UK estimates  \cite{Ferguson.2020} with about 35-fold overload without interventions (UK baseline, upper dotted line) and no overload with UK lockdown (lower dotted line) enforcing school and university closure, case isolation and general social distancing. Note the logarithmic vertical axes in all panels to make vast case number differences visible simultaneously with the capacity.

\newpage

{
\centering
\includegraphics[width=\textwidth]{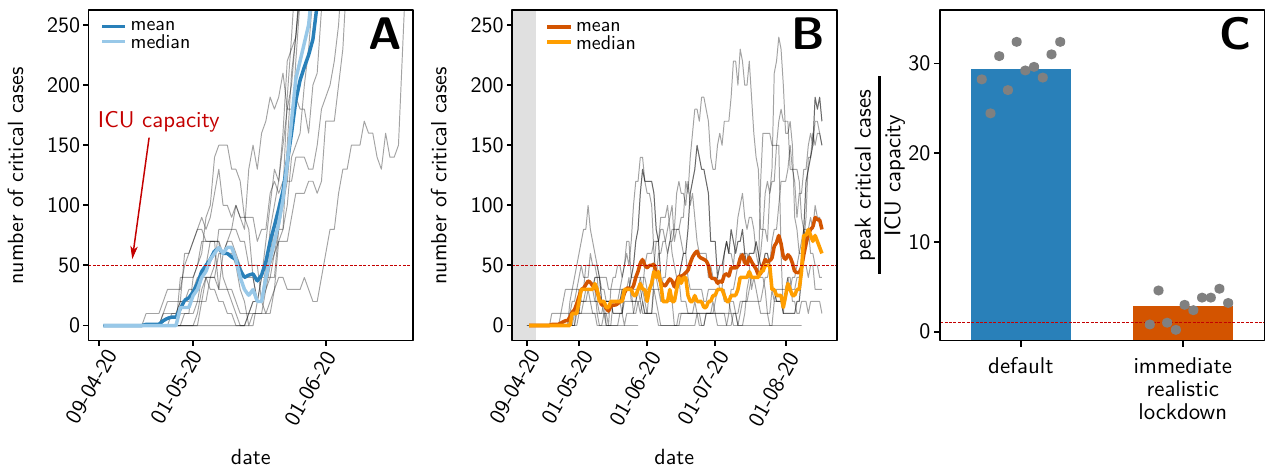}
}
\noindent {\bf Fig. 3. Confining the impact by rapid lockdown.} Ten stochastic realizations of agent-based simulations (gray lines in (A) and (B)) show increasing typical number of critical cases, quantified by the ensemble mean (dark) and median (light). 
(A) Stochastic dynamics of the number of critical cases when keeping the currently active default measures in NMBM. The ICU capacity is exceeded in almost all realizations within two months. (B) Switching from \textit{default} (currently roughly enacted in South Africa) to the \textit{realistic lockdown} scenario, limiting all activity by $90\%$, four days after crossing 100 infected people (estimated from 30 actually reported cases) in NMBM (that is on April 13, 2020), yields an exceeded ICU capacity in 80\% of realizations, at most five times above the available capacity.
(C) Relative exceedance of available ICU capacity in both scenarios. Keeping a default would cause roughly 3000\% overload whereas enacting a stricter lockdown on April 13th substantially reduces expected overload to less than 300\% and may completely avoid an overload in best cases.

\clearpage

\renewcommand{\thefigure}{S\arabic{figure}}
\renewcommand{\thetable}{S\arabic{table}}
\renewcommand{\theequation}{S\arabic{equation}}

\setcounter{figure}{0}
\setcounter{table}{0}
\setcounter{equation}{0}

\setcounter{page}{1}

{\centering
{\Large{Supplementary Material}}\\
accompanying the manuscript\\
{\Large{\center{Limited containment options of COVID-19 outbreak\\ revealed by regional agent-based simulations\\ for South Africa\\}}}
}

\vspace{1cm}
\normalsize 
\noindent Andreas Bossert\,\textsuperscript{1,7}, 
Moritz Kersting\,\textsuperscript{2,7}, 
Marc Timme\,\textsuperscript{3,6}$^\ast$, 
Malte Schr{\"o}der\,\textsuperscript{3}, 
Azza Feki\,\textsuperscript{4,7}, 
Justin Coetzee\textsuperscript{5},
and Jan Schl{\"u}ter\,\textsuperscript {6,7}$^\ast$\\

\vspace{1cm}

\noindent \normalsize{$^{1}$Center of Methods in Social Sciences, Department of Social Sciences, Georg August University of Göttingen, Goßlerstraße 19, 37073 Göttingen, Germany}\\[0.3em]
\normalsize{$^{2}$Chair of Statistics, Department of Economics, Georg August University of Göttingen, Humboldtallee 3, 37073 Göttingen, Germany}\\[0.3em]
\normalsize{$^{3}$Chair for Network Dynamics, Institute for Theoretical Physics and Center for Advancing Electronics Dresden (cfaed), Technical University of Dresden, Helmholtzstr. 18, 01069 Dresden, Germany}\\[0.3em]
\normalsize{$^{4}$Chair of Software Engineering, Faculty of Natural Sciences and Technology, Hochschule für angewandte Wissenschaft und Kunst, Von-
Ossietzky-Straße 99, 37085 Göttingen, Germany}\\[0.3em]
\normalsize{$^{5}$GoMetro, 10 Church Street, Durbanville, Cape Town, South Africa, 7550}\\[0.3em]
\normalsize{$^{6}$Institute for the Dynamics of Complex Systems, Faculty of Physics, Georg August University of Göttingen, Friedrich-Hund-Platz 1, 37077 Göttingen, Germany}\\[0.3em]
\normalsize{$^{7}$NGM, Department of Dynamics of Complex Fluids, Max-Planck-Institute for Dynamics and Self-Organization, Am Fassberg 17, 37077 G\"ottingen, Germany}

\newpage

\section*{Materials and Methods}

\subsection*{Transport simulation}

For the transport simulation, we employ the latest version (version 12.0-SNAPSHOT) of MATSim \cite{Horni.2016}. The analysis relies on the population data file provided by Joubert \cite{JOUBERT20181012}. It processes travel diaries from the 2004 Travel Survey to compute a synthetic population sample of the Nelson Mandela Bay Municipality (NMBM). Table \ref{tab:descriptive} outlines the descriptive statistics of the synthetic agents and the modal split. It becomes apparent that the minibus taxis are the backbone of the region’s transport system. %besides walking, [not a part of the "transport system"]
Private transport such as passenger cars play a rather marginal role. This difference underlines the interdependence of geographical, historical and economic characteristics of the region. 

\begin{table}[!htbp]
\centering
\begin{tabular}{lp{1cm}r} 
\toprule
\textit{Variables} & & \textit{Values} \\ 
\hline
Number of Agents ($10\,\%$ Sample) & & 114.346 \\
Number of minibus taxis (DRT) & & 2300 \\
 & &  \\
\textit{Number of Legs with mode:} & & \\
\hline
Walk & & 208.694 \\
Minibus taxi & & 104.730\\
Car (driver)& & 20.257 \\
Car (non-driving passenger) & & 21.803\\
Train & & 265 \\
Bike & & 158 \\
\bottomrule
\end{tabular}
\caption{\textbf{Descriptive Statistics of the Nelson Mandela Bay Municipality population sample.} The 114.346 agents represent a $10\%$ sample of the total population %of about 1.15 Million inhabitants 
of NMBM. The trip distribution by transport mode shows the high prevalence of low cost transport modes (walk and minibus taxi) compared to private car usage. (Data from  \cite{JOUBERT20181012,neumann2015}))
} 
\label{tab:descriptive}
\end{table}

We employ the Demand Responsive Transport (DRT) framework for MATSim \cite{bischoff2017} to include the informal minibus taxi transit. In the underpinning MATSim, agents walk to a bus stop and request a DRT vehicle (in this case a minibus taxi). Their ride is pooled with rides of other agents with similar destinations. The simulated routing is more flexible than in reality, since the minibus taxis operate on a stop-based system with routes. Due to a lack of detailed data, this mode was implemented based on a door-to-door based operating scheme and pick ups waiting passengers from the stops. 
Consequently the vehicles are less frequented on the one hand but travel longer distances with passengers on board to pick up customers in the city on the other. This reduces the likelihood of being on a minibus with an infected person, but increases the contact time if an infected person is actually on board. The formal bus transport services in NMBM are provided by Algoa Bus Company \cite{algoabus}. Due to limited data on frequencies and schedules, the formerly bus passengers are assigned to the minibus taxis in the simulation.

\subsection*{Transport parameters}

Due to the 10\,\% population sample, the total capacity of the minibus taxis vehicles must be adapted to reflect the proportions in reality. In 2014, a total of 2,374 minibus taxis operated in NMBM with an average capacity of 15 persons \cite{neumann2015}. As scaling the capacity of minibus taxis to 1.5 passengers would strongly underestimate the infections during minibus taxi trips, several test scenarios with different fleet sizes and passenger capacities were carried out.

Due to the operating scheme of door-to-door DRT a reduction either in fleet size and/or passenger capacity of the vehicles would lead to a high rejection rate and accordingly high infection numbers at home facilities that could bias the simulation results. For this reason, the number of vehicles has been chosen with respect to the trade-off between capacity utilisation and the rejection rate. A fleet of 2,300 vehicles with a capacity of 15 optimises these criteria and was introduced into the model. The vehicles were placed randomly in the area at the beginning of the simulation, although in reality they wait for customers at designated places at the beginning of the day and start their tour when a certain degree of occupancy is reached.

Public minibus taxis are assumed to take an important role in the epidemic simulation, as both the probability and the intensity of contact are assumed to be high. Moreover, people with different places of work and residence mix up at this small space.

\subsection*{Epidemic simulation}

The model relies on the MATSim-based Episim-framework to simulate the epidemic spreading in the research area. The following briefly summarises both the functionality of the default Episim configuration and the parametric adaptions to the NMBN. It is important to note that the package is still in a very early stage of development. For the following simulations, we used the latest version (master in GitHub) dated April 11, 2020 (see also https://github.com/matsim-org/matsim-episim).\\

Episim is based on a traditional SIR model, which is a common model for the analysis of epidemics \cite{Kermack.1927} and has been continuously improved, e.g. by \cite{Keeling.2008, Anderson.2010, Roche.2011}. The basic mechanism is that people go through different stages during an epidemic, and have different characteristics with each transition. In short, initially all persons are susceptible for a disease and, over time, become infected with a given probability, partly influenced by individual characteristics. Later on, they recover. %with another certain probability. 
The states and transitions are usually extended to a more complex framework and include quarantine, seriously sick and critical patients in order to account for policy measures and the need for either hospital beds or ICUs. \\

The infection process is based on a probabilistic model and occurs in ``containers''. These containers represent locations where several agents may interact, such as households, workplaces or transport vehicles, and are computed based on the information from NMBM MATSim simulation output event data. These chronicle all trajectories covered by agents during the day and the vehicles and facilities they visited and stayed at. Once a susceptible and a contagious agent stay in the same container, an infection occurs with a certain probability, which is described by equation \ref{eq:InfectionEquation} (see below).

\subsection*{Epidemic parameters}

At the beginning of the simulation, 10 randomly selected agents (corresponding to 100 people) are initially infected. After 7 days of uncontrolled spread, various countermeasures are implemented. We consider the dynamics until the countermeasures are in effect for 60 days. \\

In Episim, infected agents undergo several state transitions before their infection is terminated. Since not all humans suffer equally from COVID-19, the agents follow different paths from infection to recovery. In the beginning of the simulation, all agents except the "Patients Zero" are initially healthy and thus \textit{susceptible}. After an infection, an agent's state changes to \textit{infected but not contagious}, due to incubation time. With the beginning of the fourth day after an infection, the agent's state changes to \textit{contagious}. From this step on, the agent's differ in their behaviour. In the default scenario, 20~\% of the agents put themselves into quarantine. The idea of the simulation model is that only a certain share of those suffering from COVID-19 notices symptoms. In addition, it is to be expected that some infected persons will go into public despite symptoms. The self-quarantine lasts 14 days and is assumed to mitigate social contacts completely, even within the household. 4.5\,\%\ of all infected agents become \textit{seriously sick} on day 10 and of these, 25~\% become \textit{critical} the following day. In this way, a distinction is made between patients who require regular medical care and those who are dependent on ICUs. Contagious agents recover 16 days after infection, whereas the infection of patients with severe conditions terminates after 23 days. \\

In reality, an infection would end by recovery or death. For usability reasons, every infected agent recovers in Episim. As recovered agents are assumed to be immune and no longer contagious, omitting death does not bias the further course. The same mechanism is applied for self-quarantined, who remain mobile in the simulation but are neither contagious nor susceptible. \\

The probability of agent $n$ becoming infected at time $t$ when leaving a container is given by
\begin{align}
    P_{n,t} &= 1- \exp \lbrack-\theta  \sum_{m} q_{m,t} i_{nm,t}  \tau_{nm,t}\rbrack \,,
    \label{eq:InfectionEquation}
\end{align}
%where P denotes the infection probability of agent $n$ at time $t$, with 
where $q$ denotes the shedding rate (infectivity-parameter for the virus), $i$ the contact intensity and $\tau$ the interaction duration of two persons, summed over all persons in contact with agent $n$. $i$ is assumed to scale as $d^{-3}$, thus declining very fast with increasing distance. $\theta$ is introduced as a calibration parameter to shape the infection curve in a realistic way. Based on current infection data a tenfold increase of infected persons in 7 days is quite realistic for most countries without any interventions. In our framework, we thus set $\theta_{NMBM} = 0.000003$ in order to meet this condition.\\

Most of the contact intensities were left at their default since it is reasonable to assume, that the contact intensities of most activities do not differ significantly from the values determined by \cite{mueller2020}. The activities \textit{work}, \textit{shopping}, \textit{dropby} and \textit{other} were introduced so that the first two have the same contact intensity as \textit{leisure} (5) and the other two take the values 7 and 3. It is assumed that due to the available space and seating arrangements in the minibus taxis, the contact between passengers in the minibus taxis is much higher than in formal public transport in Berlin. Since no reliable data on the contact intensity of passengers are available, we take the interaction intensity to be 20, compared to 10 in the default setting. Likewise, the contact intensity of being at home is doubled from 3 to 6 due to limited space, larger households and general living conditions.\\

\subsection*{Policy Parameters}
The countermeasures in the default setting, realistic lockdown and complete lockdown are described by modifiers to the activity rates of the agents, describing the reduction in the frequency of the respective activity. Table~S2 list all parameters for the scenarios studied.

\begin{table}[!htbp]
\centering
\begin{tabular}{lp{1cm}rrrr}
\hline

Scenario &              & Baseline & Default & Realistic & Lockdown   \\
\hline
&&&&&\\
Home &                  & 1  & 1    & 1    &  1     \\
Work  &                 & 1  & 0.2  & 0.1  &  0        \\
Minibus taxis  &        & 1  & 0.7  & 0    &  0         \\
Leisure &               & 1  & 0    & 0.1  &  0         \\
KiGa and Prim. Educ. &  & 1  & 0    & 0    &  0           \\
Higher Education  &     & 1  & 0    & 0    &  0          \\
Shopping  &             & 1  & 0.3  & 0.1  &  0        \\
Dropby  &               & 1  & 0    & 0    &  0          \\
Other  &                & 1  & 0.2  & 0.1  &  0         \\
&&&&&\\
\hline
\end{tabular}
\caption{\textbf{Allowed share per activity in 4 scenarios.} Episim allows to prohibit or restrict the practice of certain activities in order to simulate social distancing. Four sets of restrictions were implemented in the simulations. While in the baseline scenarios no measures are in power, a complete lockdown does not allow the population to leave their homes. Between those extreme cases, two different partial lockdowns are set up.} 
\label{tab:policies}

\end{table}

\newpage

\section*{Simulation results}
Here we provide additional measures recorded from the simulations presented in the main manuscript on the location of infection events (Table~S3) and exact values of the health care system overload (Table~S4).

%\subsection*{Identifying infection sites}

% \ \\

\begin{table}[!htbp]
\centering
\begin{tabular}{lp{1cm}rrrr}
\hline

Scenario &              & Baseline & Default & Realistic & Lockdown   \\
\hline
&&&&&\\
\multicolumn{3}{l}{\textit{Scenarios with Minibus infection }}  \\
Home &                  & 75,259  & 39,629 & 9,063 &  1,774     \\
Work  &                 & 2,188   & 919    & 171  &  5        \\
Minibus taxis  &        & 16,002  & 3,705  & 799  &  18         \\
Leisure &               & 3,105   & 524    & 206  &  0         \\
KiGa and Prim. Educ. &  & 13,560  & 2,745  & 690  &  1           \\
Higher Education  &     & 579     & 119    & 41   &  0          \\
Shopping  &             & 915     & 549    & 26   &  0        \\
Dropby  &             & 22      & 3      & 1    &  0          \\
Other  &                & 2,210   & 621    & 99   &  4         \\

&&&&&\\

\multicolumn{3}{l}{\textit{Scenarios without Minibus infection }}        \\
Home &                  & 83,747  & 17,421  & 3,417  & 1,091          \\
Work &                  & 2,589   & 670    & 45    & 5           \\  
Minibus taxis &         & 0      & 0      & 0     & 0              \\
Leisure &               & 4,075   & 392    & 91    & 0           \\
KiGa and Prim. Educ.&   & 16,581  & 2,068   & 142   & 2           \\
Higher Education &      & 822    & 74     & 8     & 0           \\
Shopping  &             & 1,257   & 288    & 7     & 0       \\
Dropby  &             & 28     & 1      & 0     & 0           \\
Other  &                & 3,021   & 398    & 32    & 4           \\
&&&&&\\
\hline

\end{tabular}
\caption{\textbf{Total number of infection events per location in the simulation scenarios.} 
Each event represents an initial infection of one out of 114.346 total agents. While most infections occur at home due to prolonged and close contact, paratransit in terms of minibus taxis is the second largest driver of the spreading process, accounting for approximately 15\% of all infection events in the baseline scenario. These numbers decrease under countermeasures and almost disappear under complete lockdown, where only a few people are initially infected before all travel is stopped. A comparison to artificial settings where infections during minibus transit are neglected demonstrate the importance of the minibus taxi service as a driver for the spreading, in particular under partial lockdown.} 
\label{tab:places}

\end{table}

\begin{table}[!htbp]
\centering
\begin{tabular}{lp{1cm}rrrr}
\hline

    Scenario    &              &  Baseline &  Default & Realistic & Lockdown \\
\hline
&&&&&\\
\textit{Scenarios with Minibus Infection}    &&&&&\\
 Peak Newly Infected per Day     &                   & 128,870 & 15,510 & 2,370  &  2,350        \\
 Peak Seriously Sick        &                  & 36,890  & 5,170  & 1,090  &  570           \\
 Peak Critical      &                          & 10,280  & 990    & 260    &  170           \\
 Peak Critical Cases per ICU       &        & 205.6   & 19,8   & 5.2    &  3.4             \\
 $\lambda$ &                                   & 118,320 & 16,965 & 4,055  &  1,100            \\
  Days until first ICU Overload &              &   18    &  21    & 20     &  20               \\
  
&&&&&\\

\textit{Scenarios without Minibus Infection}    &&&&& \\
 Peak Newly Infected per Day     &                   &  109,400 & 8,500  & 1,930  &  1,920      \\
 Peak Seriously Sick        &                  & 34,720   & 2820   & 400    &  370        \\
 Peak Critical      &                          & 9,030    & 750    & 130    &  120        \\
 Peak Critical Cases per ICU  &             & 180.6    & 15     & 2.6    &  2.4        \\
 $\lambda$ &                                   & 113,515  & 6,020  & 1,055  &  475        \\
  Days until first ICU Overload &              & 20       & 26     & 24     &  23         \\
&&&&&\\

\hline
\end{tabular}
\caption{\textbf{Exact values of health care system overload measures} for all simulation scenarios. During the first 6 days, the virus spreads uncontrolled. On day seven, countermeasures are initiated that restricts certain activities (see \ref{tab:policies}). The upper values result from simulations in which minibus taxis are possible locations of infection (presented in the main manuscript). The lower values are drawn from simulations, in which people could not become infected in minibus taxis (compare Table~S3).
}
\label{tab:key_figures}

\end{table}

%\begin{table}[!htbp]
%\centering
%\begin{tabular}{lp{1cm}rrr}
%\hline
%
%
%  Start of severe lockdown       &          & Day 4 &  Day 7 & Day 14 \\
%\hline
%&&&&\\
%  Min Peak Critical  &                      &  10 &   10 &     20    \\
% Mean Peak Critical &                       &  138 &   149 &    108   \\
% Max Peak Critical &                        &  240 &   220 &       210      \\
% Fraction of Simulations with ICU overload &   & 8/10 & 9/10 & 9/10      \\
%  Mean Days until first ICU Overload &      &   33 &   48 &         28.7    \\
%  (conditional on overload) & &  &  & \\
%  
%  &&&& \\
%\hline
%
%\end{tabular}
%\caption{\textbf{Healthcare system impacts of a delayed severe lockdown} in the realistic setting illustrated in Fig.~3 in the main manuscript. Here, the default countermeasures are in effect from day 1, corresponding to 09-04-20. Depending on the scenario, this comparably soft policy is replaced by a severe realistic lockdown on day 4, 7 or 14, corresponding to  13-04-20, 16-04-20 and 23-04-20, respectively.} 
%\label{tab:healthcare_overload_trend}
%\end{table}

%{
%\centering
%\includegraphics[width=\textwidth]{pics/grid_at_samstag/grid_day7_ohne_bus.pdf}
%}
%
%\noindent {\bf Fig. S1. Containment scenarios of COVID-19} in the Nelson Mandela Bay Municipality (NMBM) with total population of 1.14 million people. Same parameters as in figure 1, except infections are not possible in paratransit (compare Table~S3).

\end{document}